\documentclass[aps,prl,twocolumn,showpacs,preprintnumbers,amsmath,amssymb,superscriptaddress]{revtex4}%

\usepackage{graphicx}
\usepackage{xspace}
\usepackage{siunitx}
\usepackage[usenames]{color}
\usepackage{comment}
\usepackage{bigdelim}
\usepackage{amssymb}
\usepackage{amsmath}
\usepackage{array}
\usepackage{multirow}
\usepackage{tabularx}
\usepackage{dcolumn}
\usepackage{booktabs,dcolumn}
\usepackage{bm}
\usepackage{float}
\usepackage{fancyhdr}
\usepackage{afterpage}
\usepackage{url}
\usepackage{bm}
\usepackage{epstopdf}

\usepackage{enumitem}
\usepackage{setspace}
\usepackage{color}
\usepackage{hyperref}
\hypersetup{colorlinks,urlcolor=blue,citecolor=black}
\usepackage{listings}
\usepackage[dvipsnames]{xcolor}

\graphicspath{{./figs/}}

\newcommand{\nba}[1]{} 

\usepackage{graphicx}
\usepackage{dcolumn}
\usepackage{bm}
\usepackage{color}
\usepackage{amsmath}
\usepackage[normalem]{ulem}

\begin{document}

\makeatletter
\@ifundefined{textcolor}{}
{%
 \definecolor{BLACK}{gray}{0}
 \definecolor{WHITE}{gray}{1}
 \definecolor{RED}{rgb}{1,0,0}
 \definecolor{GREEN}{rgb}{0,1,0}
 \definecolor{BLUE}{rgb}{0,0,1}
 \definecolor{CYAN}{cmyk}{1,0,0,0}
 \definecolor{MAGENTA}{cmyk}{0,1,0,0}
 \definecolor{YELLOW}{cmyk}{0,0,1,0}
}


\makeatother

\preprint{preprint(\today)}

\title{Magnetism in Semiconducting Molybdenum Dichalcogenides}


\author{Z.~Guguchia}
\email{zg2268@columbia.edu}\affiliation{Department of Physics, Columbia University, New York, NY 10027, USA}
\affiliation{Laboratory for Muon Spin Spectroscopy, Paul Scherrer Institute, CH-5232 Villigen PSI, Switzerland}

\author{ A.~Kerelsky}
\thanks{These authors contributed equally to this work.}
\affiliation{Department of Physics, Columbia University, New York, NY 10027, USA}

\author{ D.~Edelberg}
\thanks{These authors contributed equally to this work.}
\affiliation{Department of Physics, Columbia University, New York, NY 10027, USA}

\author{S. Banerjee}
\affiliation{Department of Applied Physics and Applied Mathematics, Columbia University, New York, NY 10027, USA}

\author{F. von~Rohr}
\affiliation{Department of Chemistry, University of Zurich, Winterthurerstrasse 190, CH-8057 Zurich, Switzerland}

\author{D.~Scullion}
\affiliation{School of Mathematics and Physics, Queen's University Belfast, UK}

\author{M.~Augustin}
\affiliation{School of Mathematics and Physics, Queen's University Belfast, UK}

\author{M.~Scully}
\affiliation{School of Mathematics and Physics, Queen's University Belfast, UK}

\author{ D.A.~Rhodes}
\affiliation{Department of Mechanical Engineering, Columbia University, New York, NY 10027, USA}

\author{Z.~Shermadini}
\affiliation{Laboratory for Muon Spin Spectroscopy, Paul Scherrer Institute, CH-5232
Villigen PSI, Switzerland}

\author{H.~Luetkens}
\affiliation{Laboratory for Muon Spin Spectroscopy, Paul Scherrer Institute, CH-5232
Villigen PSI, Switzerland}

\author{A.~Shengelaya}
\affiliation{Department of Physics, Tbilisi State University, Chavchavadze 3, GE-0128 Tbilisi, Georgia}
\affiliation{Andronikashvili Institute of Physics of I.Javakhishvili Tbilisi State University,
Tamarashvili str. 6, 0177 Tbilisi, Georgia}

\author{C.~Baines}
\affiliation{Laboratory for Muon Spin Spectroscopy, Paul Scherrer Institute, CH-5232
Villigen PSI, Switzerland}

\author{E.~Morenzoni}
\affiliation{Laboratory for Muon Spin Spectroscopy, Paul Scherrer Institute, CH-5232
Villigen PSI, Switzerland}

\author{A.~Amato}
\affiliation{Laboratory for Muon Spin Spectroscopy, Paul Scherrer Institute, CH-5232
Villigen PSI, Switzerland}

\author{ J.C.~Hone}
\affiliation{Department of Mechanical Engineering, Columbia University, New York, NY 10027, USA}

\author{R.~Khasanov}
\affiliation{Laboratory for Muon Spin Spectroscopy, Paul Scherrer Institute, CH-5232
Villigen PSI, Switzerland}

\author{S.J.L.~Billinge}
\affiliation{Department of Applied Physics and Applied Mathematics, Columbia University, New York, NY 10027, USA}
\affiliation{Condensed Matter Physics and Materials Science Department,
Brookhaven National Laboratory, Upton, NY 11973, USA}

\author{E.~Santos}
\email{e.santos@qub.ac.uk} \affiliation{School of Mathematics and Physics, Queen's University Belfast, UK}

\author{ A.N.~Pasupathy}
\email{apn2108@columbia.edu} \affiliation{Department of Physics, Columbia University, New York, NY 10027, USA}

\author{Y.J.~Uemura}
\email{tomo@lorentz.phys.columbia.edu} \affiliation{Department of Physics, Columbia University, New York, NY 10027, USA}


\maketitle
\textbf{Transition metal dichalcogenides (TMDs) are interesting for understanding fundamental physics of two-dimensional materials (2D) as well as for many emerging technologies, including spin electronics \cite{Soluyanov, Xu, Ali1, ZhuZ,  PanXC, QiCava, Moncton, Wilson,Qian, Morosan, LiY}.
Here, we report the discovery of long-range magnetic order 
below $T_{M}=$40 K and 100 K in bulk semiconducting TMDs 2H-MoTe$_{2}$ and 
2H-MoSe$_{2}$, respectively, by means of muon spin-rotation (${\mu}$SR), scanning tunneling microscopy (STM), as well as density functional theory (DFT) calculations. The muon spin rotation measurements show the presence of a large and homogeneous internal magnetic fields at low temperatures in both compounds indicative of long-range magnetic order. DFT calculations show that this magnetism is promoted by the presence of defects in the crystal. The STM measurements show that the vast majority of defects in these materials are metal vacancies and chalcogen-metal antisites which are randomly distributed in the lattice at the sub-percent level. DFT indicates that the antisite defects are magnetic 
with a magnetic moment in the range of 0.9-2.8 $\mu_B$. Further, we find that the magnetic order stabilized in 
2H-MoTe$_{2}$ and 2H-MoSe$_{2}$ is highly sensitive to hydrostatic pressure. 
These observations establish 2H-MoTe$_{2}$ and 2H-MoSe$_{2}$ as a new class of magnetic semiconductors and opens a path to studying the interplay of 2D physics and magnetism in these interesting semiconductors.}

\begin{figure*}[t!]
\includegraphics[width=0.98\linewidth]{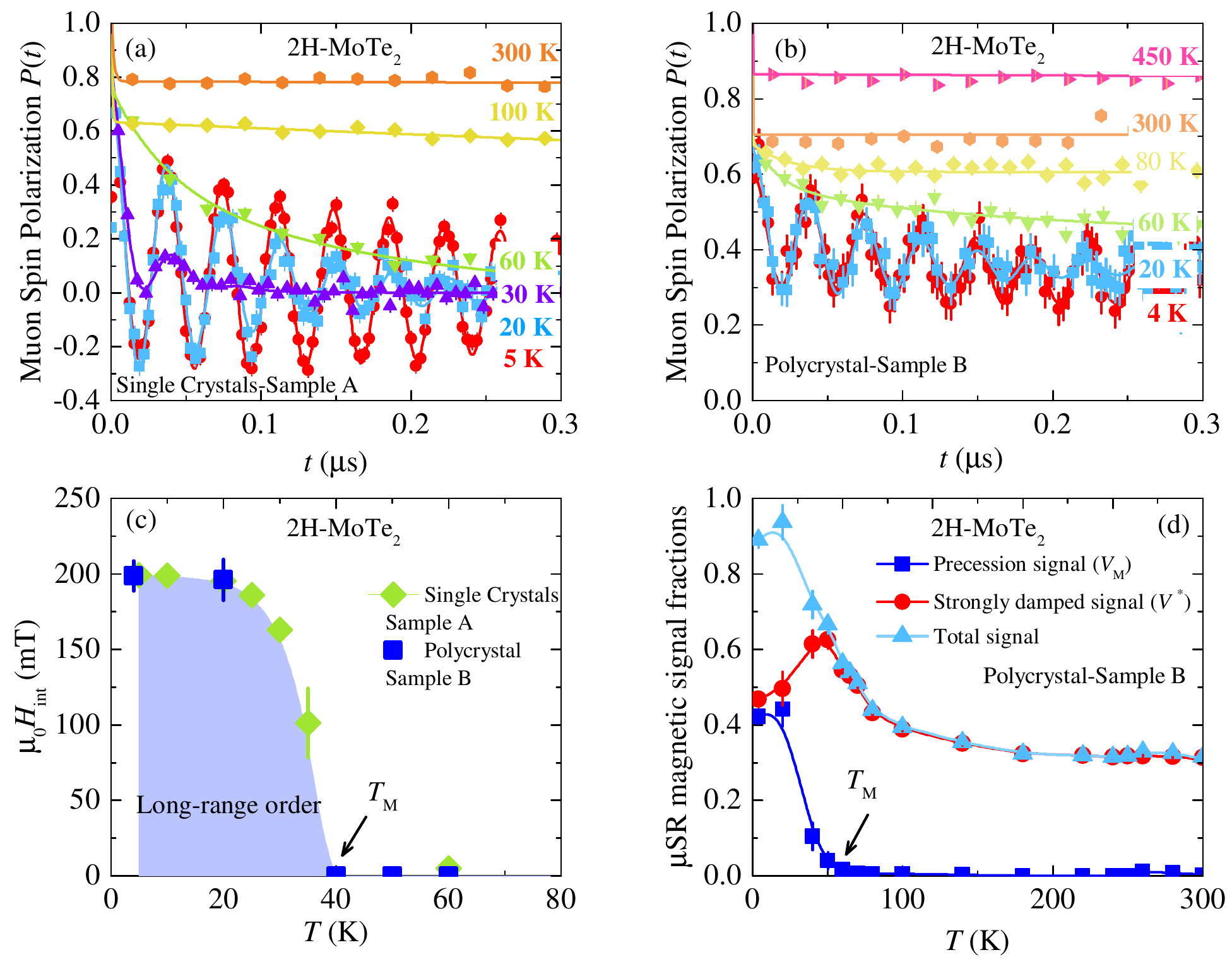}
\vspace{-0.4cm}
\caption{(Color online) \textbf{ZF ${\mu}$SR time spectra and temperature dependent ZF ${\mu}$SR parameters for MoTe$_{2}$.}
ZF ${\mu}$SR time spectra for the single crystal (a) and polycrystalline (b) samples of MoTe$_{2}$ recorded at various temperatures up to 450 K. (c) The temperature dependence of the internal field ${\mu}_{0}H_{int}$ of 2H-MoTe$_{2}$ as a function of temperature. (d) The temperature dependence of  the magnetic fractions $V_{\rm M}$ and $V^{*}$ of the precession and strongly damped signals, respectively (see text). The total magnetic signal is also shown.} 
\label{fig7}
\end{figure*}

 Transition metal dichalcogenides, a family of 2D layered materials like graphene, have been subject of a tremendous amounts of experimental and theoretical studies due to their exciting electronic and optoelectronic properties  \cite{Soluyanov, Xu, Ali1, ZhuZ,  PanXC, QiCava, Moncton, Wilson, Qian, Morosan, LiY}. TMDs share the same formula, MX$_{2}$, where M is a transition metal and X is a chalcogen. They have a layered structure and crystallize in several polytypes, including 2H-, 1T-, 1T$^{'}$- and $T_{d}$-type lattices \cite{Clarke,Puotinen,Brown}. 
Much interest has focused on the cases of M= Mo or W, since the 2H forms of these compounds are semiconducting and can be mechanically exfoliated to a monolayer. In bulk form, 2H-MoTe$_{2}$ is a semiconductor with an indirect band gap of 0.88 eV.
The unique properties of TMDs especially in the monolayer form have shown great promise in device applications such as:
magnetoresistance and spintronics, high on/off ratio transistors, optoelectronics, valley-optoelectronics, superconductors and hydrogen storage \cite{Moncton, Wilson, Qian, Morosan, LiY}. Many of these interesting properties arise on account of the strong spin-orbit interaction present in these materials due to the heavy metal ion. While there are many studies focused on the spin-orbit coupling and the interesting consequences for electrical and optical properties in these systems, there are very limited, and mostly theoretical, studies on intrinsic magnetism in these materials \cite{Li, Ataca, Terrones, Shidpour, Ma, Yan, Tongay, Zheng,Wei2015}. Theoretical and experimental work shows that in the absence of crystalline imperfections, the Mo-based TMDs are nonmagnetic \cite{Ataca}. The ability to magnetism into the properties of these materials can open up a host of new opportunities as tunable magnetic semiconductors.

 In this letter, we report muon spin relaxation/rotation (${\mu}$SR) and scanning tunneling microscopy (STM) experiments carried out on both polycrystalline and single crystalline samples of 2H-MoTe$_{2}$ and 2H-MoSe$_{2}$, as well as Hubbard-corrected DFT calculations (DFT+$U$) to gain insights into the experimental results.  ${\mu}$SR experiments serve as an extremely sensitive local probe technique to detect small internal magnetic fields and ordered magnetic volume fractions in magnetic materials. STM has the ability to measure atomic and electronic structure with atomic resolution, and has been employed extensively in the past to study local electronic properties in TMDs and other 2D materials \cite{kerelsky}.The techniques of STM and ${\mu}$SR complement each other ideally as we are able to study the magnetic properties of these crystals sensitively with ${\mu}$SR experiments, and correlate these magnetic properties with atomic and electronic structure measured by STM. Experimental details are provided in the Methods Section.

\begin{figure*}[t!]
\centering
\includegraphics[width=1.0\linewidth]{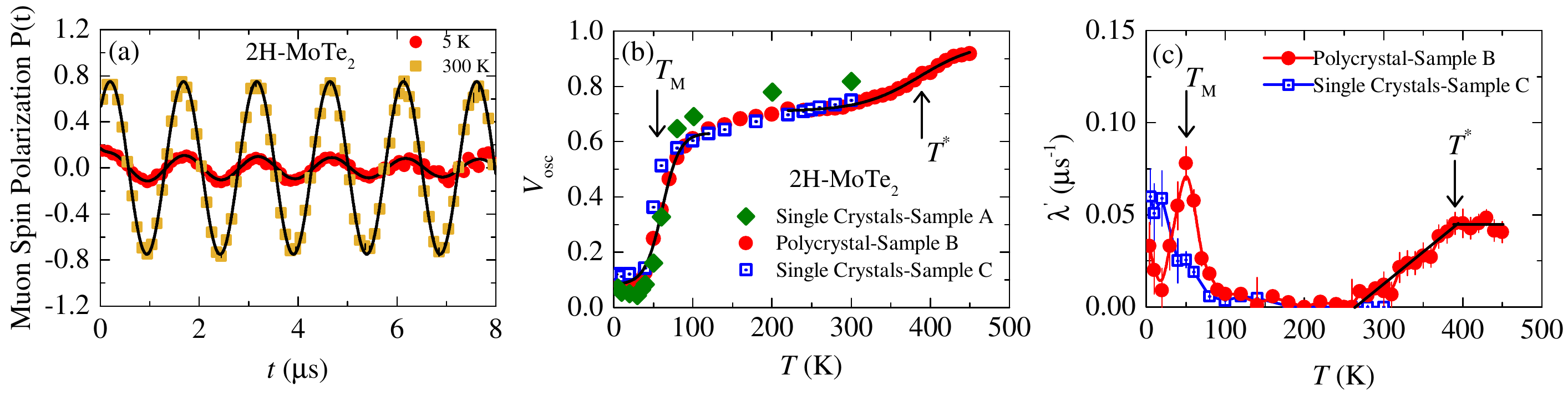}
\vspace{-0.8cm}
\caption{ (Color online) \textbf{Temperature dependent weak-TF ${\mu}$SR parameters and weak-TF ${\mu}$SR spectra  for MoTe$_{2}$.} (a) WTF ${\mu}$SR time spectra for MoTe$_{2}$ recorded at $T$ = 5 K and 300 K. The solid gray lines represent fits to the data by means of Eq. (2). (b) The bulk magnetic moment of the polycrystalline sample of MoTe$_{2}$ measured in an applied magnetic field of $H$ = 10 mT. The temperature dependence of the oscillating fraction (c) and the paramagnetic relaxation rate ${\lambda}$ (d) of the single crystalline and polycrystalline samples of MoTe$_{2}$ obtained from the weak-TF ${\mu}$SR experiments. The solid arrows mark the magnetic transition temperatures $T_{M}$ and $T^{*}$. The solid gray lines represent fits to the data by means of phenomenological function (see Eq. (3) of the Method section).}
\label{fig4}
\end{figure*}

 Zero-field ${\mu}$SR time spectra for single crystalline (Sample A) and polycrystalline (Sample B) samples of MoTe$_{2}$, recorded for various temperatures in the range between 4 K and 450 K, are shown in Figs. 1a and b, respectively. At the highest temperature $T$ = 450 K (Fig. 1b), nearly the whole sample is in the paramagnetic state. The paramagnetic state causes only a very weak depolarization of the ${\mu}$SR signal. This weak depolarization and its Gaussian functional form are typical for a paramagnetic material and reflect the occurrence of a small Gaussian Kubo-Toyabe depolarization, originating from the interaction of the muon spin with randomly oriented nuclear magnetic moments. Upon cooling, first a fast decaying ${\mu}$SR signal is found. Below $T_{\rm M}$ ${\simeq}$ 40 K, in addition to the strongly damped signal,  a spontaneous muon-spin precession with a well-defined frequency
is observed, which is clearly visible in the raw data (Fig. 1a and b). Figure 1c shows the temperature dependence of the local magnetic field ${\mu}_{0}H_{int}$ at the muon site for both single crystalline and polycrystalline samples of MoTe$_{2}$. There is a smooth increase of ${\mu}_{0}H_{int}$ below  $T_{\rm M}$ ${\simeq}$ 40 K, reaching the saturated value of ${\mu}_{0}H_{int}$ = 200 mT at low temperatures. Observation of the spontaneous muon-spin precession indicates the occurrence of long range static magnetic order in semiconducting 2H-MoTe$_{2}$ below $T_{\rm M}$ ${\simeq}$ 40 K, a remarkable finding. It is important to note that, the long-range magnetic order was also observed in single crystalline and polycrystalline  samples of 2H-MoSe$_{2}$ (see the supplementary Figure 6), but with higher ordering temperature $T_{\rm M}$ ${\simeq}$ 100 K and with the higher local magnetic field ${\mu}_{0}H_{int}$ ${\simeq}$ 300 mT. This difference might be related to the different magnetic structures in these two samples 2H-MoSe$_{2}$ than in 2H-MoTe$_{2}$. Figure 1d exhibits the temperature dependence of the ${\mu}$SR signal fractions (oscillating and strongly damped) in the polycrystalline sample of MoTe$_{2}$. At the base $T$ = 4 K, oscillations are observed for about 45 ${\%}$ of the muons while about 45 ${\%}$  show a strong relaxation. Additionally, it is evident from the ZF ${\mu}$SR data that  the oscillating component develops at the cost of the strongly damped fraction, since the appearance of the oscillating component below $T$ = 40 K is accompanied  by the reduction of the  strongly damped fraction (see Fig. 1d). While the presence of the oscillating signal in the zero-field ${\mu}$SR time spectra is an unambiguous signature of long-range magnetism in these compounds, the strongly damped signal observed in these compounds can arise for a few different reasons. The simplest possibility is that it is related the presence a strongly disordered magnetic phase in the sample with a  broad static field distribution. It is equally possible that it arises from fluctuating magnetic moments that  might exist at non-zero temperature, especially above the actual long-range magnetic order temperature $T_{\rm M}$ ${\simeq}$ 40 K. A third possibility is the presence of some muonium fraction in semiconducting 2H-MoTe$_{2}$ which would also gives rise to the strongly damped signal. Muonium a bound state of ${\mu}^{+}$ and an electron may form in semiconductors. In the bound state the muon is much more sensitive to magnetic fields than as a free probe, since its magnetic moment couples to the much larger electron magnetic moment thus amplifying the depolarization effects. Therefore even small variations in the magnetic field may cause a strong depolarization such as that observed in the spectra at early times.

\begin{figure*}[t!]
\centering
\includegraphics[width=1.0\linewidth]{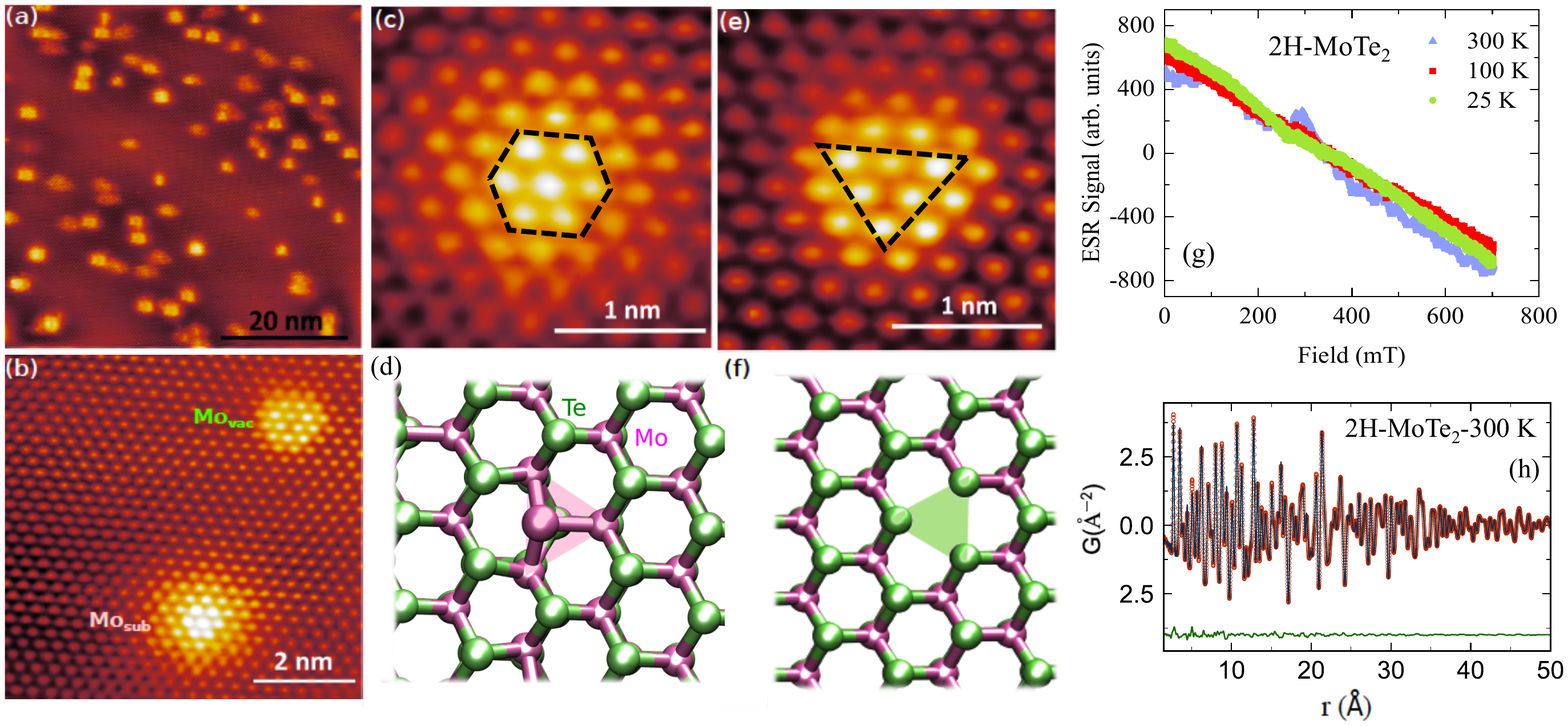}
\vspace{-4.1cm}
\caption{\textbf{Observation of intrinsic defects in 2H-MoTe$_{2}$ through scanning 
tunneling microscope (STM) and sample characterizations.} (a) Large scale 
atomic resolution STM topography (20 nm) of the MoTe$_{2}$ surface. The image reveals an approximately 
uniform density of two types of defects over the entire surface. The STM topography was taken at -1.25 V and -100 pA setpoint. 
(b) Small scale atomic resolution STM topography (2 nm) show that these two type of defects 
are mainly substitutional Mo atoms at Te sites (Mo$_{sub}$) and Mo vacancies (Mo$_{vac}$). 
(c)-(d) Local STM topography (1 nm) and DFT$+U$ optimized geometry for Mo$_{sub}$ defect, respectively. 
The observed atoms in (c) are those at the top layer of tellurium, with an increased 
topographic height profile at the center of the six brightest spots. We attribute this to 
a molybdenum replacement of a tellurium atom. 
(e)-(f) Local scale STM topography (1 nm) and DFT$+U$ optimized geometry
of the second type of defects observed, respectively. The image in (e) shows a 
depression in the topographic height profile, centered between three tellurium atoms. 
Based on the symmetry, we attribute this to a molybdenum vacancy under 
the layer of tellurium. (g) Electron Spin Resonance spectra for 2H-MoTe$_{2}$, recorded at various temperatures. (h) Pair Distribution Function average structure refinements for 2H-MoTe$_{2}$ at 300 K fitted to the hexagonal 2H-structure model.}
\label{fig4}
\end{figure*}

  In order to obtain precise information about the ordered magnetic volume fraction of MoTe$_{2}$, weak-transverse field (TF) ${\mu}$SR experiments were carried out. In weak-TF experiments, the amplitude of the low-frequency oscillations of the muons precessing in the applied field is proportional to the nonmagnetically-ordered volume fraction. Thus, a spectrum with no oscillation corresponds to a fully ordered sample, while a spectrum with oscillation in the full asymmetry indicates a non-magnetic sample. The weak-TF spectra for MoTe$_{2}$, recorded at $T$ = 4 K and 300 K are shown in Fig. 2a, which show intermediate oscillation amplitudes, indicating that over a broad temperature interval, this material contains both magnetic and non-magnetic regions. This compound therefore exhibits intrinsic non-magnetic and magnetic phase separation. Figure 2b displays the fraction of the low-frequency oscillations $V_{\rm osc}$ = 1 - $A_{S}^{'}$($T$)/$A_{S}(0)$ (see the analysis section) as a function of temperature in two different single crystalline (4 K ${\leq}$ $T$ ${\leq}$ 450 K) as well as in the polycrystalline samples of MoTe$_{2}$ (4 K ${\leq}$ $T$ ${\leq}$ 300 K). At 450 K,  $V_{\rm osc}$ exhibits nearly a maximum value, indicating that the whole sample is in the non-magnetic state, and all the muon spins precess in the applied magnetic field. $V_{\rm osc}$ decreases with decreasing temperature below 425 K and tends to saturate below 300 K. This changes below 100 K, where an additional substantial  decrease of $V_{\rm osc}$ is observed to value $V_{\rm osc}$ ${\simeq}$ 0.1 below 40 K. This implies that below 40 K, only 10 ${\%}$ of the sample is non-magnetic. 
The temperature dependence of the fraction $V_{\rm osc}$ is in fair agreement with the total ${\mu}$SR signal fraction, obtained from the ZF ${\mu}$SR experiments (Fig. 1d).  The relaxation rate ${\lambda}^{'}$ of the paramagnetic part of the signal (see the analysis section for the details) also shows significant features in its temperature dependence. There is a clear peak at $T_{\rm M}$ ${\simeq}$ 40 K, which is a signature of a magnetic phase transition. In the large temperature range beyond the peak ${\lambda}^{'}$ is constant and starts to increase above 250 K, reaching its maximum value at $T^{*}$ ${\simeq}$ 400 K. This supports the presence of some transition in the sample at $T^{*}$ ${\simeq}$ 400 K. It is important to emphasise that the weak-TF data obtained for two single crystal samples agree very well with each other. Moreover, the data for the polycrystalline MoTe$_{2}$ sample is also in excellent agreement with the single crystal data. This implies that the observed transitions  and long range magnetic ordering below 40K are reproducible and real. The combination of the ZF and TF ${\mu}$SR experiments allow us to conclude that there are two phases in the system  2H-MoTe$_{2}$: (a) A Low-$T$ phase which is characterized by the long-range static magnetic order ($T_{\rm M}$ ${\simeq}$ 40 K) and (b)  a second phase, which appears below $T^{*}$ ${\simeq}$ 400 K. This high-$T$ transition can arise for few different reasons and it is most likely not magnetic in origin. 
 
\begin{figure*}[t!]
\centering
\includegraphics[width=1.0\linewidth]{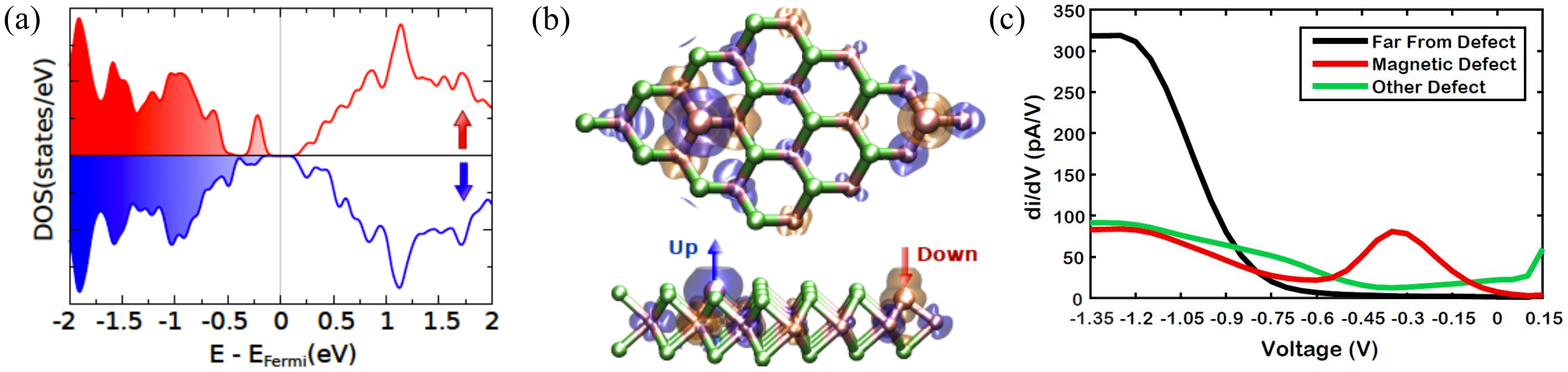}
\vspace{-8.3cm}
\caption{\textbf{Hubbard-corrected density functional theory calculations (DFT+$U$) and STM.} 
(a) Spin polarized density of states, DOS(states/eV), of Mo$_{sub}$ defects in the antiferromagnetic (AFM) phase. 
Fermi level, E$_{Fermi}$, is set to zero. Both the spin up and down DOS reveal an in-gap state due to the defect. (b) Magnetization density ($\pm$0.001 $electrons/Bohr^3$) on top surface of bulk 
2H-MoTe$_{2}$ in AFM configuration. Spin up and spin down states are shown in faint blue and orange iso-surfaces, respectively. 
Note that spins also couple antiferromagnetically at local level between the Mo impurity and the nearest Mo atoms. (c) Scanning tunneling spectroscopy dI/dVs taken on the two types of defect as well as far from any defect.}
\label{fig4}
\end{figure*}

Previous theoretical work  \cite{Ataca} as well as simple chemical bonding considerations indicate that the Mo atoms in 2H-MoTe$_{2}$ are in a nonmagnetic 4d$^2$ configuration. We therefore investigate the presence of defects in the crystals measured by ${\mu}$SR and ask whether we can associate them with the observed magnetism. In order to do this, we perform atomic-resolution scanning tunneling microscopy (STM) measurements on in-situ cleaved surfaces on crystals from the same batch from which ${\mu}$SR measurements were performed.
Shown in figure 3a is a typical STM topograph of the surface of MoTe$_{2}$. The density of defects observed is low enough that they can simply be counted individually to make an estimate of the defect density in the crystal. The defect density measured in Figure 3a is 0.4 ${\%}$, a number that is typical of all samples measured. To understand these defects further, high resolution STM imaging was performed to resolve the lattice site at which defects are observed. Being a local surface probe, STM imaging resolves only the top tellurium of the MoTe$_{2}$ lattice clearly. We use the Te atoms to infer defect locations based on their relative heights and centers. Two types of defects were observed within these crystals, one located on the Te site (Fig. 3b, c and d) and the other located at the metal site (Fig. 3b, e and f). From the relative densities of these two defects we find a majority of sites to be the former Te site defect. We see from the topography of this defect that the defect is associated with a foreign atom replacing the chalcogen rather than a chalcogen vacancy. 
While STM measurements themselves cannot identify the chemical nature of the defect, we have performed Electron Spin Resonance (ESR) experiments to investigate the presence of paramagnetic impurities such as Fe or Ni. The results of these measurements are shown in
Fig. 3g (see also the Supplementary Figure 7), which indicate no trace of ESR signal down to the lowest temperature. Previous transmission electron microscopy (TEM) measurements \cite{kerelsky} have also shown that one of the prominent defect types in these materials is a chalcogen antisite, where a molybdenum atom replaces the tellurium atom. We therefore proceed to identify it as such. The other type of defect (Fig. 3e and f), which is only present in low density, is at the site of the metal atom. Based also on previous TEM measurements of defects, we identify this with a Mo vacancy (Mo$_{vac}$) in the crystal. This identification is also consistent with STM images which show that the vacancy is a topographic depression at all biases measured. Finally, average structure Pair Distribution Function (PDF) refinement for 2H-MoTe$_2$ (Fig. 3h) (see also the Supplementary Figure 8) confirms the 2H polytype (SG:$P6_{3}/mmc$) \cite{puotinen_crystal_1961} and shows no evidence of structural distortions or segregation, consistent with the dilute concentration of intrinsic defects. This is in line with the observation of a random distribution of defects from STM.

 Having identified the primary defect types in our crystals, we perform Hubbard-corrected DFT simulations (DFT+$U$, see {\it Methods} for details) to examine their magnetic properties. In the absence of the Hubbard $U$, the defects are found to be non-magnetic. At finite values 
of $U$, a magnetic moment in the range of 0.9-2.8 $\mu_B$ is observed per Mo antisite impurity. Along with magnetism, the calculations find that in the presence of $U$, there are small distortions from triangular symmetry at the Mo$_{sub}$ defect site. The spin-polarized density of states (Fig. 4a) shows that the localized Mo $4d$-states at the Fermi level carry most of the magnetization with minor contribution from $p$-states of the Te atoms. We also find that he Mo$_{sub}$ defects are coupled antiferromagnetically to the nearest neighbor Mo atoms as shown in Fig. 4b. The magnetic moments at the nearest neighbor Mo atoms can reach 0.10-0.40 $\mu_B$/atom, with smaller contributions for second- and third-neighbors (0.02-0.08 $\mu_B$/atom). The Te atoms show negligible spin-polarization. Similar effects have previously been observed in graphene with different adsorbates and substitutional metal atoms \cite{Santos10a,Santos10b,Santos12}. The metal vacancy Mo$_{vac}$ does not introduce a significant local moment in our calculations. Shown in Figure 4c are STM tunneling spectra taken on the two types of defect as well as far from any defect. Spectra taken on the Mo$_{sub}$ defects always display a deep in-gap state, while the metal vacancy Mo$_{vac}$ does not show such a feature. Comparing to the DFT calculation in Fig. 4a, we see that this in-gap state is consistent with the DFT model for a molybdenum replacement of a tellurium. 
   
\begin{figure}[t!]
\centering
\includegraphics[width=1.0\linewidth]{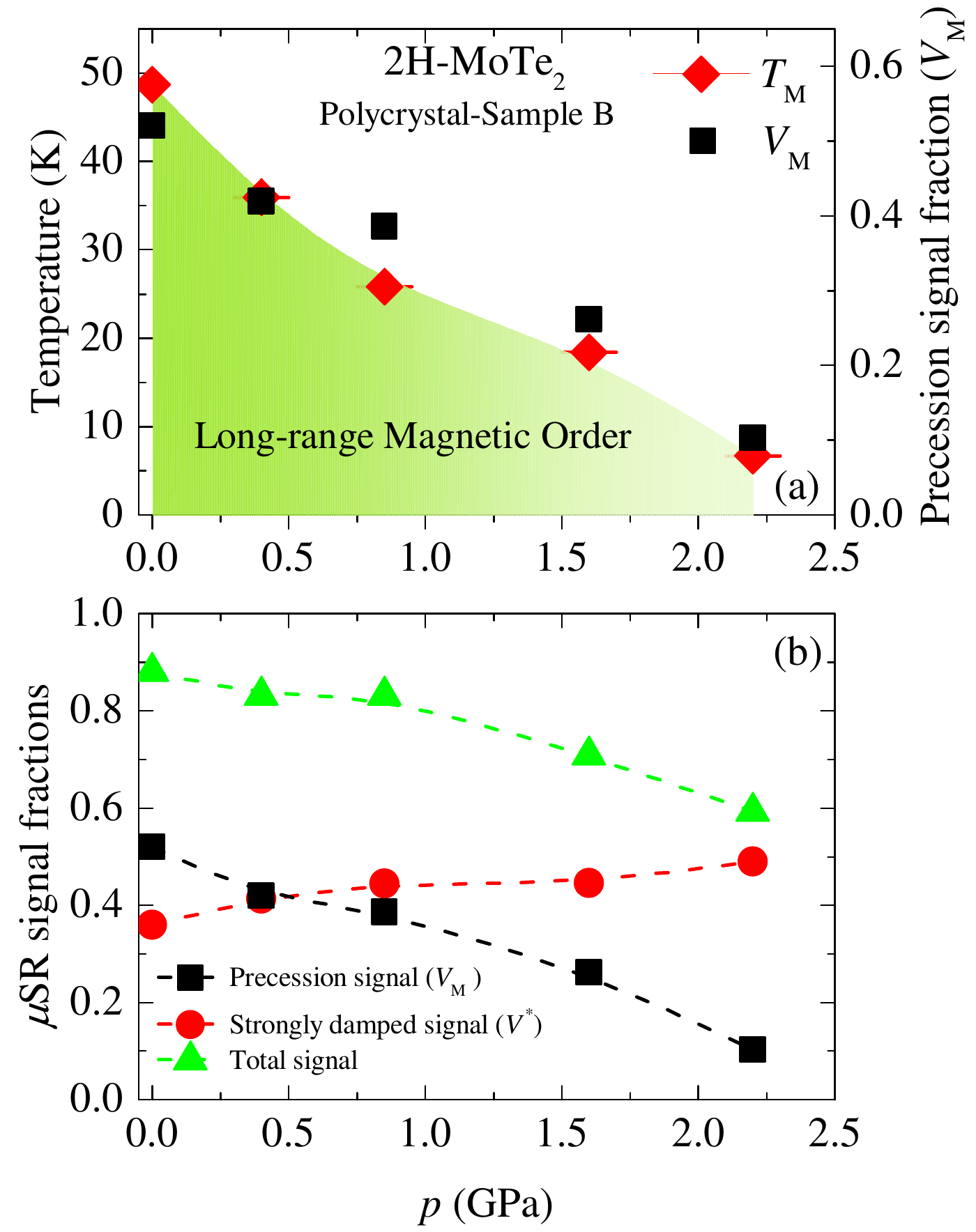}
\vspace{-0.7cm}
\caption{ (Color online)  \textbf{Pressure evolution of various quantities.} (a) The magnetic transition temperature $T_{\rm M}$, the SC transition temperature $T_{\rm c}$, and the magnetic volume fraction $V_{\rm M}$ as a function of pressure. (b) The pressure dependence of the magnetic fractions 
$V_{\rm M}$ and $V^{*}$, corresponding to the precession ${\mu}$SR signal and the strongly damped ${\mu}$SR signal, respectively. The total magnetic signal is also shown. The dashed lines are the guides to the eyes.}
\label{fig4}
\end{figure}

 For  further insight into the magnetic order in MoTe$_{2}$, weak-TF ${\mu}$SR experiments were carried as a function of hydrostatic pressure. 
We indeed find that hydrostatic pressure has a significant effect on the magnetic properties of these materials.
From the high-pressure ${\mu}$SR spectra (see the Supplementary Figure 9), we can construct a temperature-pressure phase diagram for 2H-MoTe$_{2}$. Shown in figure 5a and 5b are the pressure dependences of the magnetic transition temperature  $T_{\rm M}$ as well as the magnetic volume fractions $V_{\rm M}$ and $V^{*}$ ($T$ = 100 K). Interestingly, we clearly see a suppression of the magnetic order as a function of pressure, which can be observed as the precession signal decreases with pressure while the strongly damped signal remains relatively uniform as shown in Fig. 5b. Further, it is clear from Fig. 5a that the transition temperature $T_{\rm M}$ decreases as a function of temperature. These findings show that one can physically tune the magnetism in these materials with pressure.

 Our muon measurements unambiguously establish 2H-MoTe$_2$ and 2H-MoSe$_2$ as magnetic, moderate band gap semiconductors.
The ${\mu}$SR results indicate a very homogeneous and strong antiferromagnetic order below $T_{\rm M}$ ${\simeq}$ 40 K and 100 K, for MoTe$_{2}$ and MoSe$_{2}$, respectively.  
In the same materials, STM measurements show the presence of intrinsic dilute self-organized magnetic telurium/selenium antisite defects, a finding that is well  supported by Coulomb-corrected density functional theory. Although, the exact link between ${\mu}$SR and STM/DFT results is not clear yet, both results together constitute first strong evidence for involvement of magnetic order in physics of TMDs. Establishing long range magnetic order with the observed low density of antisite defects would necessarily involve electronic coupling to the semiconductor valence electrons. The presence of such spin-polarized itinerant electrons implies that these materials are dilute magnetic semiconductors (DMS). Previously DMS materials have been synthesized in a range of thin film \cite{Ohno} and crystal \cite{CuiDing} materials. Much interest has centered on the III-V semiconductor class \cite{Ohno,Jungwirth,Dietl,Dunsiger}, where a small concentration of some magnetic ions, particularly Mn$^{2+}$, can be incorporated by substituting for the group-III cations of the host semiconductor.  Numerous technical challenges in making uniform DMS materials have been overcome in recent years \cite{Tseng,Saadaoui,Parker,Zhao2013,CuiDing,ChenBJ}, but formidable challenges still remain in producing stable, high-quality DMS materials with high $T_{c}$. Our present system offers unique advantages to these other routes to synthesize DMS materials. First, the defects contributing to magnetism are intrinsic in the crystal, and are uniformly distributed. This can alleviate some of the materials challenges commonly faced in DMS synthesis. Second, the materials are cleavable and readily grown in large area form down to a monolayer thickness \cite{Cain}. As is well known in these materials, the bandgap is a strong function of thickness, giving us tunability over the semiconductor properties. Third, the chemical potential and electric field in thin films is easily tuned by electrostatic gates \cite{LiNC}, allowing for the possibility of tunable magnetism as has been seen for GaAs \cite{Ohno,Jungwirth}. Finally, these materials can be easily layered by van der Waals heteroepitaxy  \cite{Koma} allowing for the creation of unique new device concepts in the future.

\section{METHODS}
\textbf{Sample preparation}: High quality single crystals and polycrystalline samples were obtained by mixing of molybdenum foil (99.95 ${\%}$) and tellurium lumps (99.999+${\%}$) in a ratio of 1:20 in a quartz tube and sealed under vacuum. The reagents were heated to 1000$^{\rm o}$C within 10 h. They dwelled at this temperature for 24 h, before they were cooled to 800$^{\rm o}$C within 30 h (polycrystalline sample) or 100 h (single crystals). At 800$^{\rm o}$C the tellurium flux was spined-off and the samples were quenched in air. The obtained MoTe$_{2}$ samples were annealed at 400$^{\rm o}$C for 12 h to remove any residual tellurium. 

\textbf{Pressure cell}:  Pressures up to 2.2 GPa were generated in a double wall piston-cylinder
type of cell made of CuBe and MP35N materials, especially designed to perform ${\mu}$SR experiments under
pressure \cite{MusrHPI}. As a pressure transmitting medium Daphne oil was used. The pressure was measured by tracking the SC transition of a very small indium plate by AC susceptibility. The filling factor of the pressure cell was maximized. The fraction of the muons stopping in the sample was approximately 40 ${\%}$.\\

\textbf{${\mu}$SR experiment}: 
 In a ${\mu}$SR experiment nearly 100 ${\%}$ spin-polarized muons ${\mu}$$^{+}$
are implanted into the sample one at a time. The positively
charged ${\mu}$$^{+}$ thermalize at interstitial lattice sites, where they
act as magnetic microprobes. In a magnetic material the 
muon spin precesses in the local field ${\mu}_{0}H_{int}$ at the
muon site with the Larmor frequency ${\nu}_{\rm \mu}$ = ${\mu}_{0}$$\gamma_{\rm \mu}$/(2${\pi})$$H_{int}$ (muon
gyromagnetic ratio $\gamma_{\rm \mu}$/(2${\pi}$) = 135.5 MHz T$^{-1}$). 

 ${\mu}$SR experiments under pressure were performed at the ${\mu}$E1 beamline of the Paul Scherrer Institute (Villigen, Switzerland, where an intense high-energy ($p_{\mu}$ = 100 MeV/c) beam of muons is implanted in the sample through the pressure cell. The low background GPS (${\pi}$M3 beamline) \cite{GPS} and low-temperature LTF (${\pi}$M3.3) instruments were used to study the single crystalline as well as the polycrystalline samples of MoTe$_{2}$ at ambient pressure.\\

\textbf{Analysis of the ZF-${\mu}$SR data of MoTe$_{2}$}
 
 In the whole temperature range the response of the sample consists of a magnetic and a nonmagnetic contribution. At low temperatures description of the magnetic part of the ${\mu}$SR signal requires a two-component relaxation function. For temperatures below 50 K a well-defined frequency is observed for about 45 ${\%}$ of the muons while about 45 ${\%}$ show a strong relaxation due to a broad static field distribution. This situation changes above 50 K where only the strongly damped signal is observed in addition to the paramagnetic signal.  The data were analysed by the following functional form using the free software package MUSRFIT \cite{AndreasSuter}:
\begin{equation}
\begin{split}
P_S(t)=V_{M}\Bigg[{\frac{2}{3}e^{-\lambda_{T}t}\cos(\gamma_{\mu}{\mu}_{0}H_{int}t)}+\frac{1}{3}e^{-\lambda_{L}t}\Bigg] \\
+V^{*}{\frac{2}{3}e^{-\lambda_{{Fast}}t}+(1-V_{M}-V^{*})e^{-\lambda_{nm}t.}}
\label{eq1}
\end{split}
\end{equation}
Here, $V_{\rm M}$ and $V^{*}$ denote the relative magnetic fraction of the oscillating and strongly damped magnetic signals, respectively. 
${\mu}_{0}H_{int}$ is the local internal magnetic field at the muon site. 
${\lambda_T}$ and ${\lambda_L}$ are the depolarization rates representing the transversal and the longitudinal 
relaxing components of the magnetic parts of the sample.
${\lambda_{nm}}$ is the relaxation rate of the nonmagnetic part of the sample.\\

\textbf{Analysis of the weak TF-${\mu}$SR data of MoTe$_{2}$}

  A substantial fraction of the ${\mu}$SR asymmetry originates 
from muons stopping in the pressure cell surrounding the sample \cite{MusrHPI}. 
Therefore, the ${\mu}$SR data in the whole temperature range were analyzed by
decomposing the signal into a contribution of the sample and a contribution of the pressure cell.
In addition the TF-${\mu}$SR spectra were fitted in the time domain with a combination of a slowly
relaxing signal with a precession frequency corresponding to the applied field of $\mu_{0}H$ = 5 mT 
(due to muons in a paramagnetic environment) and a fast relaxing
signal due to muons precessing in much larger static local fields:
\begin{equation}
\begin{split}
A_0P(t) = A_S(0)P_S(t)+A_{PC}(0)P_{PC}(t) = \\ (A_{PC}e^{-\lambda_{PC} t}  + A_{S}^{'}e^{-\lambda^{'} t}){\cos} (\gamma_{\mu}B^{'}t)+       \\
A_{S}^{''}\Bigg[{\frac{2}{3}e^{-\lambda_{T}^{''}t}J_0(\gamma_{\mu}B^{''}t)}+\frac{1}{3}e^{-\lambda_{L}^{''}t}\Bigg],
\end{split}
\end{equation}
where $A_{\rm 0}$ is the initial asymmetry, i.e., the amplitude of the oscillation in the fully paramagnetic state. $P(t)$ is the muon spin-polarization function, and $\gamma_{\mu}/(2{\pi}) \simeq 135.5$~MHz/T is the muon gyromagnetic ratio. $A_{PC}$ and ${\lambda}_{PC}$ are the asymmetry and the relaxation rate of the pressure cell signal. $A_{S}^{'}$ and $A_{S}^{''}$ are the amplitudes of the slowly (paramagnetic) and fast relaxing sample signals, respectively.  ${\lambda}^{'}$ is the relaxation rate of the paramagnetic part of the sample, caused by the paramagnetic spin fluctuations and/or nuclear dipolar moments. ${\lambda}_{T}^{''}$ and ${\lambda}_{L}^{''}$ are the transverse and the longitudinal relaxation rates, respectively, of the magnetic part of the sample. $B^{'}$  and  $B^{''}$ are the magnetic fields, probed by the muons stopped in the paramagnetic and magnetic parts of the sample, respectively. From these refinements, the paramagnetic fraction at each temperature $T$ was estimated as $V_{osc}$ = 1 - $A_{S}^{'}$($T$)/$A_{S}(0)$.\\ 

\textbf{Analysis of the temperature dependence of $V_{osc}$ in MoTe$_{2}$}

   The values of $T_{\rm M}$ and $T^{*}$ were determined by using the phenomenological function \cite{GuguchiaPRL}:  
\begin{equation}
A(T)/A_{0} = a\Bigg[1-\frac{1}{{\exp}[(T-T_{X})/{\Delta}T_{X}]+1}\Bigg]+b,
\label{eq1}
\end{equation}
where $X$ = M, ${M}^{*}$. ${\Delta}T_{\rm X}$ is the width of the transition, and $a$ and $b$ are empirical parameters. 
Analyzing the data in Fig.~2 with Eq.~(3) yields:
$T_{\rm M}$ = 50(3) K and $T^{*}$ = 340(3) K.\\

\textbf{$ab$ $initio$ density functional theory methods} \\

Calculations were based on $ab$ $initio$ density functional theory using the VASP code \cite{DFT1}. The generalized gradient approximation \cite{DFT3} together with van der Waals corrected functionals \cite{DFT4,vdW-d2} were utilized. The latter guarantees if any modification of the dispersion coefficients due to the spin coupling would affect the interlayer distance. No appreciable variations were observed.  A well-converged plane-wave cutoff of 800 eV was used in all calculations. Projected augmented wave \cite{DFT5,DFT6} (PAW) potentials have been used in the description of the bonding environment for Mo, Se, and Te. Atoms and cell volumes were allowed to relax until the residual forces were below 0.0001 eV/\AA under the conjugate gradient algorithm. To model the system studied in the experiments, we created large supercells containing up to 300 atoms to simulate bulk 2H-MoTe$_{2}$ and 2H-MoSe$_{2}$ with different defects and concentrations. The Brillouin zone was sampled with a 3$\times$3$\times$2 grid under the Monkhorst-Pack scheme \cite{DFT7} to perform relaxations. Energetics and electronic density of states were calculated using a converged 11$\times$11$\times$3 k-sampling for the unit cell of 2H-MoTe$_{2}$ and 2H-MoSe$_{2}$. Calculations included a Hubbard-$U$ correction \cite{hubbard} within the range of 0.5 to 4.0 eV to account for the strong on-site interactions on the doped system, in particular, those due to the localized $d$-states at Fermi level. In addition to this we used a Fermi-Dirac distribution with an electronic temperature of $k_{B}$$T$ = 20 meV to resolve the electronic structure.

\section{Acknowledgments}~
The ${\mu}$SR experiments were carried out at the Swiss Muon Source (S${\mu}$S) Paul Scherrer Insitute, Villigen, Switzerland. X-ray PDF measurements were conducted on beamline 28-ID-2 of the National Synchrotron Light Source II, a U.S. Department of Energy (DOE) Office of Science User Facility operated for the DOE Office of Science by Brookhaven National Laboratory under Contract No. DE-SC0012704. Z. Guguchia gratefully acknowledges the financial support by the Swiss National Science Foundation (Early Postdoc Mobility SNFfellowship P2ZHP2-161980). The STM work was supported by the US National Science foundation via grant DMR-1610110 (A.K.) and by the Office of Naval Research grant number N00014-14-1-0501 (D.E.). STM equipment support is provided by the Air Force Office of Scientific Research via grant FA9550-11-1-0010 (A.N.P.). The material preparation at Princeton was supported by the Gordon and Betty Moore Foundation EPiQS initiative, grant GBMF-4412. Work at Department of Physics of Columbia University is supported by US NSF DMR-1436095 (DMREF) and NSF DMR-1610633 as well as REIMEI project of Japan Atomic Energy Agency. Work in the Billinge group was supported by U.S. Department of Energy, Office of Science, Office of Basic Energy Sciences (DOE-BES) under contract No. DE-SC00112704. S. Banerjee acknowledges support from the National Defense Science and Engineering Graduate Fellowship program. Sample synthesis (D. Rhodes) was supported by the NSF MRSEC program through Columbia in the Center for Precision Assembly of Superstratic and Superatomic Solids (DMR-1420634).  
A.S. acknowledges support from the SCOPES grant No. SCOPES IZ74Z0-160484. 
Z.G. acknowledges A.R. Wieteska for useful discussions. Z.G. acknowledges P.K. Biswas for his participation in the initial ${\mu}$SR experiments. Z.G. is grateful to I. Mazin for valuable discussions. E.J.G.S. acknowledges the use of computational resources from the UK national high performance computing service (ARCHER) for which access was obtained via the UKCP consortium and funded by EPSRC grant ref EP/K013564/1; the Extreme Science and Engineering Discovery Environment (XSEDE), supported by NSF grants number TG-DMR120049 and TG-DMR150017; and we are grateful to the UK Materials and Molecular Modelling Hub for computational resources, which is partially funded by EPSRC (EP/P020194/1). The Queen's Fellow Award through the startup grant number M8407MPH, the Enabling Fund (QUB, A5047TSL) and Department for the Economy (USI 097) are also acknowledged.
 

\newpage

\section{Supplementary Information}

\subsection{Zero-field ${\mu}$SR data for MoSe$_{2}$}

\begin{figure}[b!]
\includegraphics[width=1\linewidth]{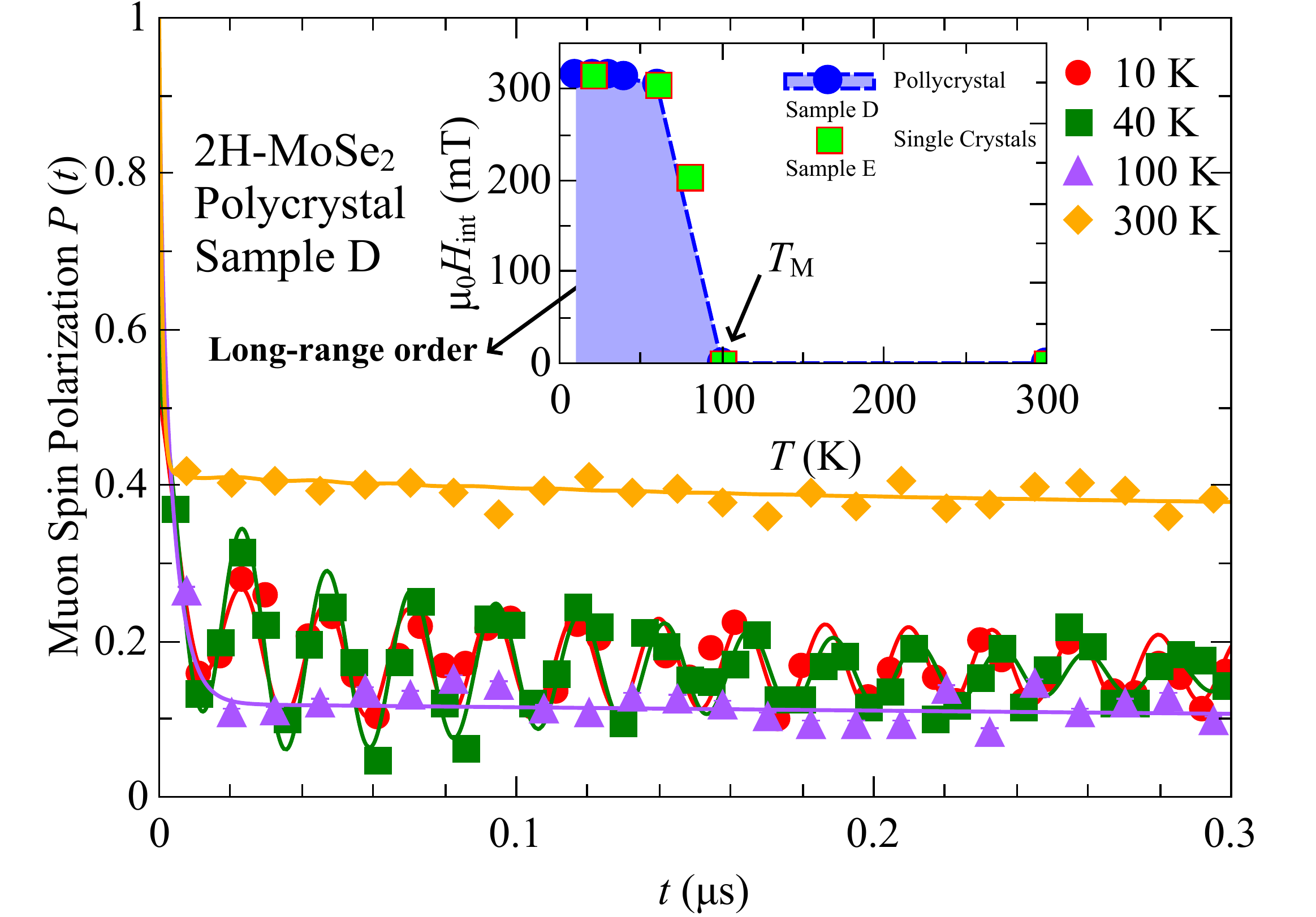}
\caption{(Color online) \textbf{ZF ${\mu}$SR time spectra and temperature dependent parameters for MoSe$_{2}$.}
ZF ${\mu}$SR time spectra for the polycrystalline samples of MoSe$_{2}$ recorded at various temperatures up to $T$ = 300 K. The inset shows the internal field $H_{\rm int}$ of MoSe$_{2}$ as a function of temperature.} 
\label{fig7}
\end{figure}

 Zero-field ${\mu}$SR time spectra for the polycrystalline (Sample D) sample of MoSe$_{2}$, recorded for various temperatures in the range between 10 K and 300 K, are shown in Fig. 6. Below $T_{\rm M}$ ${\simeq}$ 100 K, in addition to the strongly relaxed signal,  a spontaneous muon-spin precession with a well-defined frequency is observed, which is clearly visible in the raw data. The inset of Fig. 6 shows the temperature dependence of the local magnetic field ${\mu}_{0}H_{int}$ at the muon site for both single crystalline (Sample E) and polycrystalline (Sample D) samples of MoSe$_{2}$. There is a smooth increase of ${\mu}_{0}H_{int}$ below  $T_{\rm M}$ ${\simeq}$ 100 K, reaching the saturated value of ${\mu}_{0}H_{int}$ = 310 mT at low temperatures.
Observation of the spontaneous muon-spin precession indicates the occurrence of long range static magnetic order in semiconducting 2H-MoSe$_{2}$, similar as for 2H-MoTe$_{2}$. However, the magnetic ordering temperature $T_{\rm M}$ ${\simeq}$ 100 K as well as the internal field 
${\mu}_{0}H_{int}$ ${\simeq}$ 310 mT in 2H-MoSe$_{2}$ is higher as compared to the ones $T_{\rm M}$ ${\simeq}$ 40 K and ${\mu}_{0}H_{int}$ ${\simeq}$ 200 mT, observed in 2H-MoTe$_{2}$. This difference might be related to the different magnetic structures in these two samples 2H-MoSe$_{2}$ than in 2H-MoTe$_{2}$. On the other hand, the fraction of the strongly damped signal is higher in 2H-MoSe$_{2}$ than in 2H-MoTe$_{2}$.

\subsection{Electron Spin Resonance measurements for 2H-MoTe$_{2}$ and 2H-MoSe$_{2}$}

\begin{figure}[b!]
\includegraphics[width=1\linewidth]{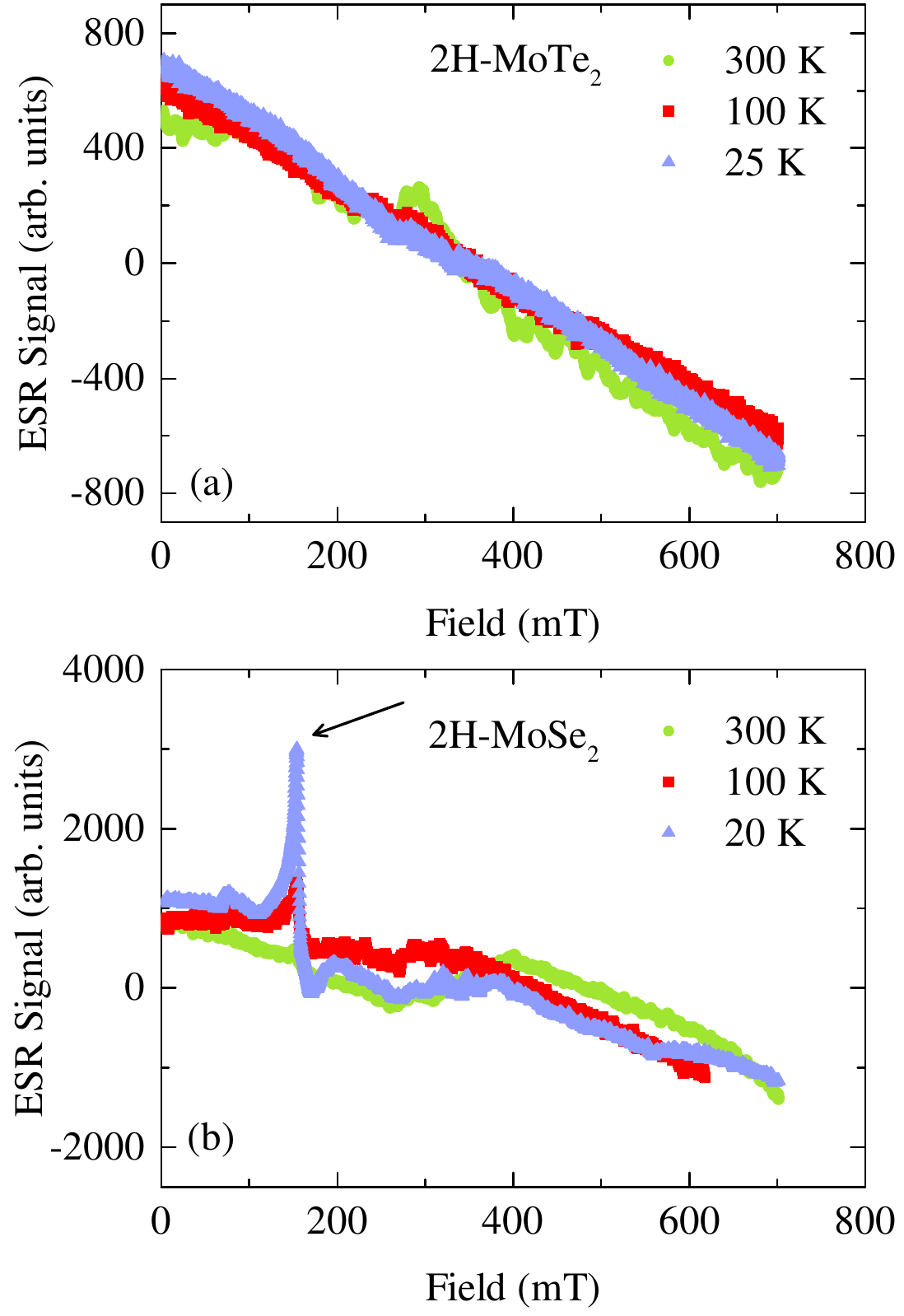}
\vspace{-0.8cm}
\caption{(Color online) \textbf{ESR signals for  2H-MoTe$_{2}$ and 2H-MoSe$_{2}$.}
Electron Spin Resonance spectra of polycrystalline samples of 2H-MoTe$_{2}$ (a) and 2H-MoSe$_{2}$ (b), taken at different temperatures. The arrow marks the ESR signal with a peak value corresponding to a $g$-factor of 4.} 
\label{fig7}
\end{figure}

\begin{figure*}[t!]
\includegraphics[width=1\linewidth]{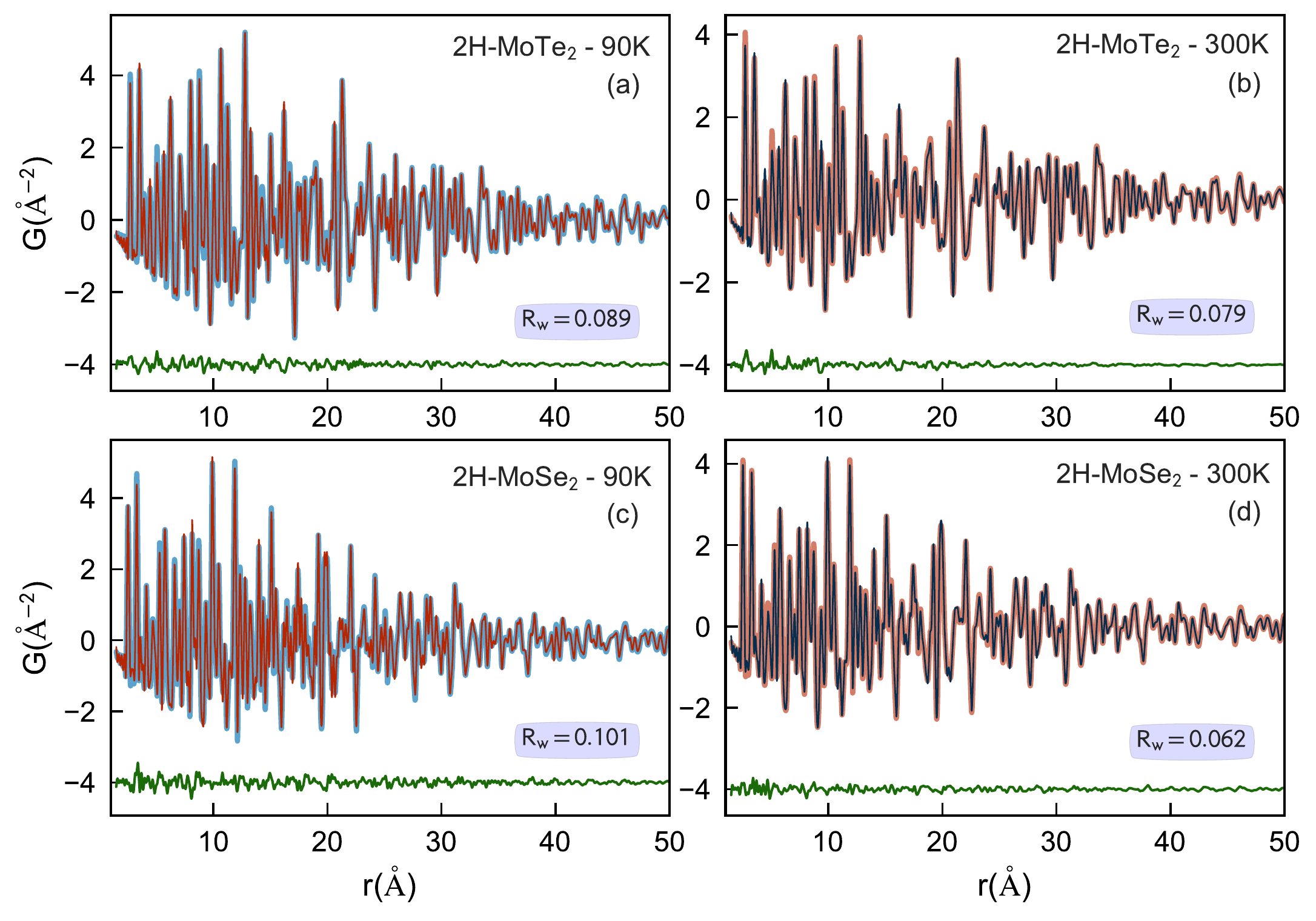}
\caption{(Color online) \textbf{PDF results for 2H-MoTe$_{2}$ and 2H-MoSe$_{2}$.}
PDF average structure refinements for 2H-MoTe$_2$ (a,b) and 2H-MoSe$_2$ (c,d) at 90~K and 300~K fit to the hexagonal 2H-structure model. Thick curves are the experimental PDFs, and the refined PDFs are overlayed as thin solid lines. Fit residuals, in green, are offset below.} 
\label{fig7}
\end{figure*}

 The polycrystalline samples of 2H-MoTe$_{2}$ and 2H-MoSe$_{2}$ have been studied by means of the Electron Spin Resonance (ESR)
technique with the emphasis on checking for the paramagnetic impurity phases.
ESR experiments were performed with a Bruker EMX spectrometer at X-band
frequencies (${\nu}$ = 9.4 GHz) equipped with a continuous He gas-flow cryostat in the temperature range 10 ${\textless}$ $T$ ${\textless}$ 300 K.
In Figure 7a and b the ESR spectra of 2H-MoTe$_{2}$ and 2H-MoSe$_{2}$, respectively, are shown for various temperatures. 
It is remarkable that no trace of ESR signal was found in the sample 2H-MoTe$_{2}$ (see Fig. 7a) down to the lowest temperature, indicating the absence of any even small amount of paramagnetic impurities such as Fe or Ni. This provides strong support to the intrinsic nature of magnetic order in  2H-MoTe$_{2}$, observed by ${\mu}$SR.
We found a small ESR signal in 2H-MoSe$_{2}$ with the $g$-factor of 4 (See Fig. 7b), which indicates the presence of small amount of paramagnetic Fe-impurities. However, this tiny amount of paramagnetic Fe-impurities would not lead to the observed long-range magnetic order in 2H-MoSe$_{2}$. ESR experiments give evidence that the magnetic order in 2H-MoTe$_{2}$ and 2H-MoSe$_{2}$ is not related to the presence of an impurity phase but it is an intrinsic state.\\

\subsection{Pair Distribution Function (PDF) Structure Confirmation of 2H-MoTe$_2$ and 2H-MoSe$_2$}

\begin{figure*}[t!]
\centering
\includegraphics[width=1.0\linewidth]{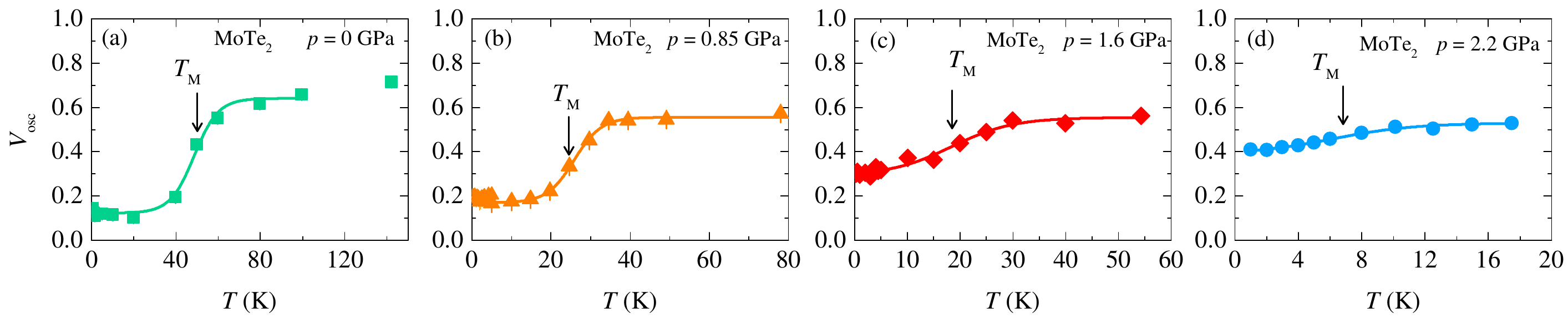}
\vspace{-0.7cm}
\caption{ (Color online)  \textbf{Temperature and pressure evolution of the paramagnetic fraction $V_{\rm osc}$.} The temperature dependence of the $V_{\rm osc}$ for the polycrystalline sample of MoTe$_{2}$ at ambient pressure (a) and at various applied pressures (b-d). The solid
lines represent fits to the data by means of Eq. (3).}
\label{fig4}
\end{figure*}

 Total scattering X-ray measurements were performed at the National Synchrotron Light Source II (XPD, 28-ID-2), Brookhaven National Laboratory. Finely ground powders of 2H-MoTe$_{2}$ and 2H-MoSe$_{2}$ were sealed in polyimide capillaries and diffraction patterns were collected in a Debye-Scherrer geometry with an X-ray energy of 66.479 keV ($\lambda=0.1865$~\AA) using a large-area 2D Perkin Elmer detector. The detector was mounted with a sample-to-detector distance of 204.14 mm. The samples were measured at 90~K and 300~K using an Oxford CS-700 cryostream. The experimental geometry, $2\theta$ range, and detector misorientations were calibrated by measuring a crystalline nickel powder directly prior to data collection at each temperature point, with the experimental geometry parameters refined using the PyFAI program~\cite{kieffer_pyfai_2013}. Standardized corrections are then made to the data to obtain the total scattering structure function, $F(Q)$, which is then Fourier transformed to obtain the PDF, using PDFgetX3~\cite{juhas_pdfgetx3:_2013} within the xPDFsuite~\cite{yang_xpdfsuite:_2014}.  The maximum range of data used in the Fourier transform ($Q_{max}$, where $Q=4\pi\sin\theta/\lambda$ is the magnitude of the momentum transfer on scattering) was chosen to be 25~\AA$^{-1}$ to give the best tradeoff between statistical noise and real-space resolution. \textsc{PDFgui} was used to construct unit cells from reference structures, carry out structure refinements, and determine the agreement between calculated PDFs and data, quantified by the residual, $R_w$ \cite{farrow_pdffit2_2007}. 

 Average structure PDF refinements for 2H-MoTe$_2$ and 2H-MoSe$_2$ are performed over a wide $r$-range from $1.5<r<50$~\AA. Lattice constants and atomic displacement parameters were constrained by hexagonal symmetry. The results of the PDF analysis are summarized in Figure 8. For both temperatures $T$ = 90~K and 300~K, the PDF is in good agreement with the 2H polytype (SG:$P6_{3}/mmc$) reported by D. Puotinen and R.E. Newnham \cite{puotinen_crystal_1961}. No evidence of structural distortions or segregation, originating from the dilute concentration of intrinsic defects was found, in line with the observation of a spatially homogeneous distribution of defects from STM. The best PDF fit for 2H-MoTe$_2$ at 300~K, yields refined lattice parameters of $a=b=3.5186$~\AA, and $c=13.9631$~\AA. For 2H-MoSe$_2$ at 300~K, $a=b=3.2875$~\AA\ and $c=12.9255$~\AA. 

\subsection{High pressure ${\mu}$SR data for MoTe$_{2}$}

 The results of the high pressure weak-TF ${\mu}$SR experiments are summarised in Figs. 9(a-d). Namely, we plot the temperature dependence of  $V_{\rm osc}$ for various hydrostatic pressures. A strong reduction of $T_{\rm M}$ as well as of the corresponding magnetic fraction is observed under pressure. For the highest applied pressure of $p$ = 1.6 GPa the $T_{M}$ decreases by  ${\sim}$ 30 K and the corresponding fraction by ${\sim}$ 50 ${\%}$.  
Note that the pressure dependence of the high temperature magnetic transition was not measured due to technical issues related to the high pressure 
${\mu}$SR technique. 

 \subsection{Magnetization measurements of MoTe$_{2}$}

   Figure 10 shows the temperature dependence of the magnetic susceptibility ${\chi}$ for MoTe$_{2}$ in an applied field of ${\mu}_{0}$$H$ = 1 T.   
In fact, semiconducting 2H-MoTe$_{2}$ is expected to exhibit temperature independent diamagnetic behavior ${\chi}_{dia}$ and a small van Vleck-type paramagnetic ${\chi}_{v}$ contribution, which is determined by the energy separation of bonding and anti-bonding states $E_{\rm a}$ - $E_{\rm b}$. 
$E_{\rm a}$ - $E_{\rm b}$ is proportional to the band gap $E_{\rm g}$. The temperature dependence of the band gap $E_{\rm g}$ of the semiconductor therefore causes a temperature dependence of ${\chi}_{v}$. Namely, ${\chi}_{v}$ is inversely proportional to $E_{\rm g}$, e.g., ${\chi}_{v}$ is expected to decrease with decreasing the temperature. As shown in Fig. 10, we observed that ${\chi}$ has tendency to decrease with decreasing temperature below 300 K, but this tendency changes at low temperatures, where ${\chi}$ increases upon lowering the temperature. Even though the magnetic response is dominated by diamagnetism, it is plausible that the total magnetic susceptibility consists of diamagnetic and paramagnetic/magnetic parts. This small paramagnetic/magnetic contribution to the total magnetization might point to the itinerant nature of magnetism in MoTe$_{2}$.\\ 

\begin{figure}[t!]
\includegraphics[width=1\linewidth]{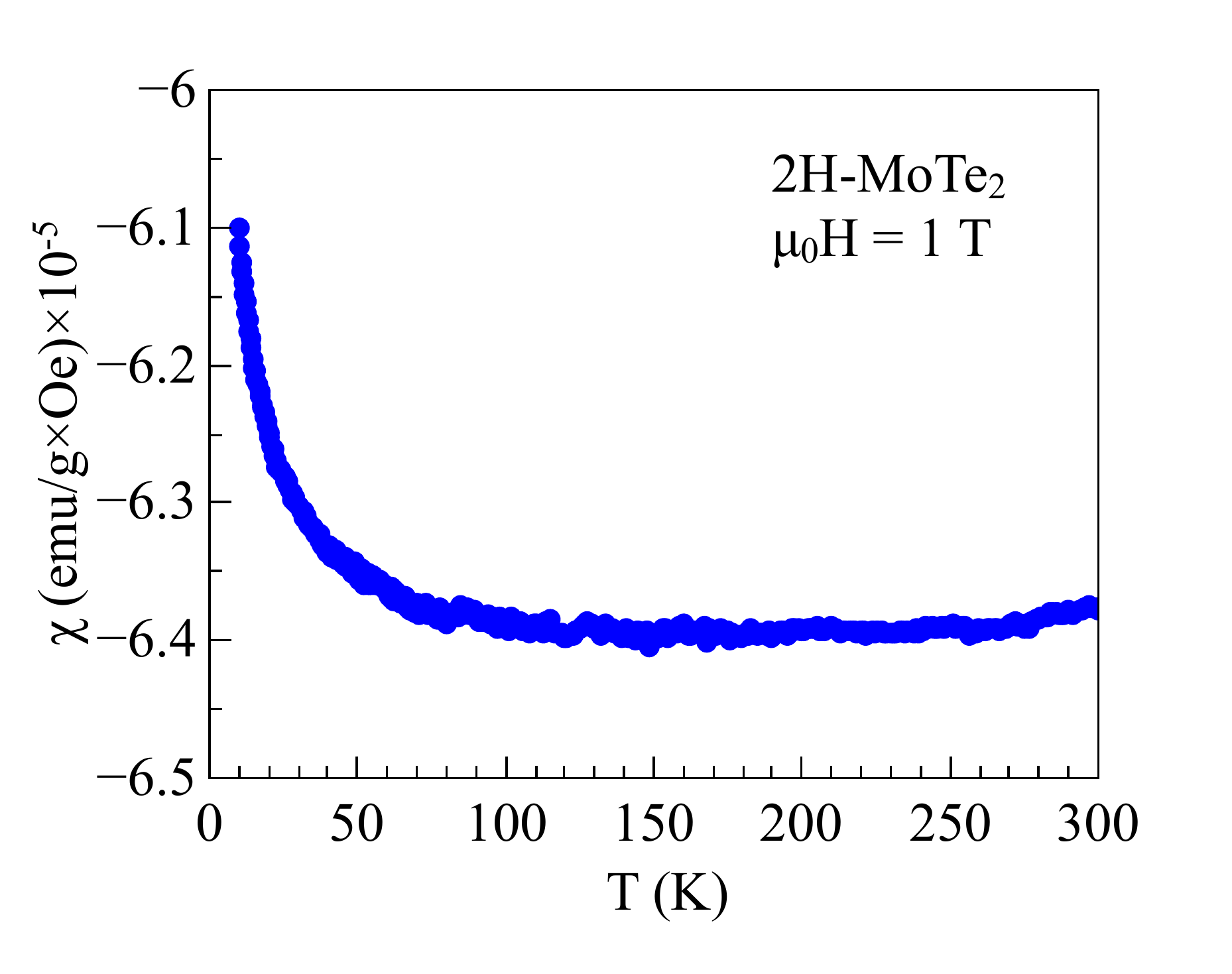}
\vspace{-1.1cm}
\caption{(Color online) \textbf{Magnetization data for MoTe$_{2}$.}
The temperature dependence of magnetic susceptibility of MoTe$_{2}$ recorded in an applied field of ${\mu}_{0}$$H$ = 1 T.} 
\label{fig7}
\end{figure}

\end{document}